\documentclass[11pt,a4paper]{article}
\usepackage[hyperref]{acl2020}
\usepackage{times}
\usepackage{latexsym}
\usepackage{graphicx}
\usepackage{url}
\usepackage{enumitem}
\usepackage{verbatim}

\aclfinalcopy

\title{\emph{Trialstreamer}: Mapping and Browsing Medical Evidence in Real-Time}

\author{
  Benjamin E. Nye \\
  Northeastern University \\
  {\small\tt nye.b@husky.neu.edu} \And
  Ani Nenkova \\
  UPenn \\
  {\small\tt nenkova@seas.upenn.edu} \AND
  Iain J. Marshall \\
  King's College London \\
  {\small\tt iain.marshall@kcl.ac.uk} \\\And
  Byron C. Wallace \\
  Northeastern University \\
  {\small\tt b.wallace@northeastern.edu}
}
\date{}

\begin{document}
\maketitle
\begin{abstract}
We introduce \emph{Trialstreamer}, a living database of clinical trial reports.
Here we mainly describe the \emph{evidence extraction} component; this extracts from biomedical abstracts key pieces of information that clinicians need when appraising the literature, and also the relations between these. Specifically, the system extracts descriptions of trial participants, the treatments compared in each arm (the \emph{interventions}), and which outcomes were measured. The system then attempts to infer which interventions were reported to work best by determining their relationship with identified trial outcome measures. In addition to summarizing individual trials, these extracted data elements allow automatic synthesis of results across many trials on the same topic. We apply the system at scale to all reports of randomized controlled trials indexed in MEDLINE, powering the automatic generation of \emph{evidence maps}, which provide a global view of the efficacy of different interventions combining data from all relevant clinical trials on a topic. 
We make all code and models freely available\footnote{https://github.com/bepnye/evidence\_extraction/} alongside a demonstration of the web interface.\footnote{http://bit.ly/trialstreamer}
% im: we need to reconcile the naming + URLs --- we have Trialstreamer already up at trialstreamer.robotreviewer.net  --- we want to integrate all this stuff, what's best way? 
% im - also ben neither link currently live, though sure you're working on it
% maybe easiest way if for now you could put links to the live system on your github page, with some text to say which is which? 
\end{abstract}

% bcw: need beautiful figures of course
% bcw: also need to contrast with ACL demo paper (/RobotReviewer) and discuss differences in aims
% bcw: consider including one of our motivating manually generated figures (from PCORI) as motivation?
\section{Introduction and Motivation}
The highest-quality evidence to inform healthcare practice comes from randomized controlled trials (RCTs). The results of the vast majority of these trials are communicated in the form of unstructured text in journal articles. Such results accumulate quickly, with over 100 articles describing RCTs published daily, on average. It is difficult for healthcare providers and patients to make sense of and keep up with this torrent of unstructured literature. 

%As an illustrative example, 
Consider a patient who has been newly diagnosed with diabetes. 
She would like to consult (in collaboration with her healthcare provider) the available evidence regarding her treatment options.
But she may not even be aware of what her treatment options are. Further, she may only care about particular outcomes (for instance, managing her blood pressure). %improving her day-to-day function and work attendance).  % bcw: i like the example (a real outcome ppl would care about) but does not align with the outcomes that seem to actually be reported! (since difficult to measure)
Currently, it is not straightforward to retrieve and browse the evidence pertaining to a given condition, and in particular to ascertain which treatments are best supported for a specific outcome of interest. 

\begin{figure}
\includegraphics[width=0.475\textwidth]{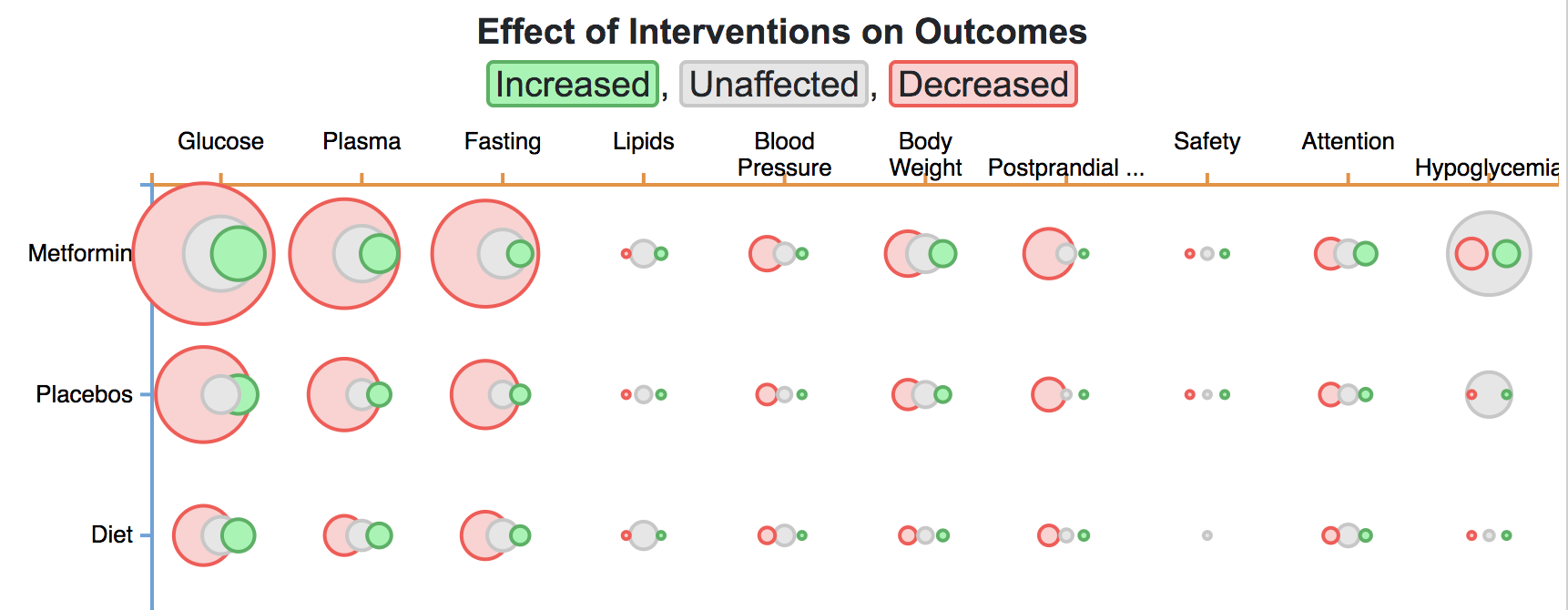}
\caption{A portion of an example evidence mapping Interventions and their inferred efficacy for Outcomes, given the condition (or Population) of \emph{Type II Diabetes}. These maps are generated automatically using the NLP system we describe in this work.}
\label{fig:ev-map}
\end{figure}

\emph{Trialstreamer} is a first attempt to solve this problem, making evidence more browseable via NLP technologies. 
Figure \ref{fig:ev-map} shows one of the key features of the system: an automatically generated \emph{evidence map} that displays treatments (vertical axis) and outcomes (horizontal) identified for a condition specified by the user (here, migraines). We elaborate on this particular example to illustrate the use of the system in Section \ref{section:example-usage}.

\begin{comment}
\begin{figure*}
\includegraphics[width=\textwidth]{figures/mouseover-conclusions.png}
\caption{Screenshot of an automatically generated evidence map relating interventions to outcomes. Bubble sizes reflect aggregate extracted trial sampled sizes, relative positioning and color communicate the inferred relative reported efficacy for the Intervention/Outcome pair. On mouse-over of individual bubbles, the system displays the article IDs used and rationales extracted from these supporting the inferred directionality of the findings; this is shown on the right.}
\label{fig:ev-map-expanded}
\end{figure*}
\end{comment}

%Healthcare providers and patients would ideally rely on rigorous \emph{evidence syntheses} of all relevant trial results to inform treatment decisions \cite{maynard1997evidence}. But performing such synthesis is laborious and time-consuming, a problem exacerbated by the rapidly growing evidence base. 

%Moreover, comprehensive, formal synthesis may not be appropriate or even desirable for all scenarios. 
%For example, one may not be able to even formulate an answerable clinical question in a particular area until becoming sufficiently familiar with the relevant literature. 
%We argue that current tools for exploring the evidence (specifically, the set of all published clinical trials) are inadequate, and we seek to address this gap by introducing a prototype tool powered by NLP that facilitates interactive evidence mapping.

Trialstreamer aims to facilitate efficient evidence mapping with a user friendly method of presenting a search across a broad field (here, being a clinical condition) \cite{Miake-Lye2016-mc}. 
We use NLP technologies to provide browseable, interactive overviews of large volumes of literature, on-demand. 
These may then inform subsequent, formal syntheses, or they may simply guide exploration of the primary literature. 
In this work we describe an open-source prototype that enables evidence mapping, using NLP to generate interactive overviews and visualizations of all RCT reports indexed by MEDLINE (and accessible via PubMed). 
When mapping the evidence one is generally interested in the following basic questions:

\begin{itemize}[leftmargin=*]
\item What interventions and outcomes have been studied for a given condition (population)? 
\item How much evidence exists, both in terms of the number of trials and the number of participants within these?
% bcw: too bad we haven't included RoB stuff yet... I mean we *have* the models though... 
% 1/27/2020 ben: taking this bullet out for now - adding RoB as future inclusion?
%\item What is the quality of this evidence? More formally, what are the \emph{risks of bias} \cite{Higgins2011}?
\item Does the evidence seem to support use of a particular intervention for a given condition?
\end{itemize}

%In the remainder of this paper we describe our prototype tool, trialstreamer, that enables interactive exploration of the evidence base -- i.e., all reports of RCTs that are available in MEDLINE -- with respect to the above questions. 
In the remainder of this paper we describe a prototype system that facilitates interactive exploration and mapping of the evidence base, with an emphasis on answering the above questions.
The Trialstreamer mapping interface allows structured search over study populations, interventions/comparators, and outcomes --- collectively referred to as PICO elements \cite{huang2006evaluation}. It then displays key clinical attributes automatically extracted from the set of retrieved trials. This is made possible via NLP modules trained on recently released corpora \cite{nye2018corpus,lehman2019}, described below. 

% bcw: not sure if we should somehow collapse the above and this one into one figure??
\begin{comment}
\begin{figure*}
\includegraphics[scale=.5]{figures/mouseover-conclusions.png}
\caption{On mouse-over of individual trials (bubbles), the system displays the rationale extracted from the abstract that supports the inferred directionality of the findings.}
\end{figure*}
\end{comment} 

\section{System Overview}

% bcw 1/28/20: watch consistency with "trialstreamer" vs "Trialstreamer" -- i don't care either way but need to be consistent.
\begin{figure*}
\includegraphics[width=\textwidth]{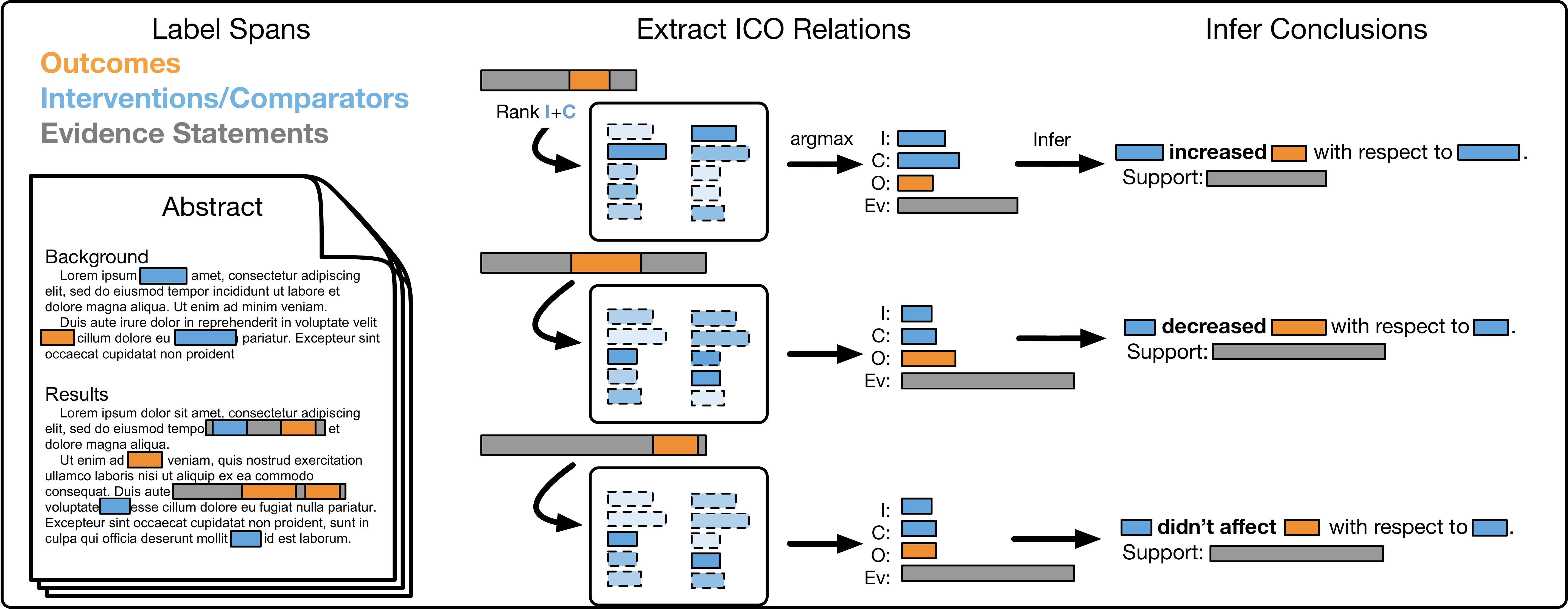}
\caption{Overview of the evidence extraction pipeline, applied to all RCT article abstracts automatically identified. Text spans are first extracted from these abstracts, then assembled into relations that reflect the structure of the trials, and finally used to infer the effect interventions were reported to have on measured outcomes, as compared to the control treatment.}
\end{figure*}

The evidence extraction pipeline is composed of four primary phases. First, text snippets that convey information about the trial's treatments (or \emph{interventions}), outcome measures, and results are extracted from abstracts. Relations between these snippets are then inferred to identify which treatments were compared against each other, and which outcomes were measured for these comparisons. The extracted relations and evidence statements are then used to infer an overall conclusion about the comparative efficacy of the trial's interventions. Finally, the clinical concepts expressed in the extracted spans are normalized to a structured vocabulary in order to ground them in an existing knowledge base and allow for aggregations across trials.

A typical RCT report would pertain to a single clinical condition (the \emph{population}), but might report multiple numerical results, each concerning a particular intervention, comparator, and outcome measure (which we describe as an ICO triplet).% bce 1/28: we didn't define "result spans" did we?

Because the end-to-end task combines NLP subtasks that are supported by different datasets, we collected new development and test sets --- 160 abstracts in all, exhaustively annotated --- in order to evaluate the overall performance of our system. Two medical doctors\footnote{Hired via Upwork (\url{http://www.upwork.com}).} annotated these documents with the all of the expressed entities, their mentions in the text, the relations between them, the conclusions reported for each ICO triplet and the sentence that contains the supporting evidence for this \cite{lehman2019}.

We were unable to obtain normalized concept labels for the ICO triplets due to the excessive difficulty of the task for the annotators.

% In contrast to more controlled biomedical categories such as chemicals or proteins, arbitrary PICO elements do not map cleanly to an ontology, often touching on several different concepts. 

Modeling decisions were informed by the 60 document development set, and we present evaluations of the first four information extraction modules with regard to the 100 documents in the unseen test set.

\subsection{Preprocessing}

Enabling search over RCT reports requires first compiling and indexing all such studies. This is, perhaps surprisingly, non-trivial. One may rely on ``Publication Type" (PT) tags that codify study designs of articles, but these are manually applied by staff at the National Library of Medicine. Consequently, there is a lag between when a new study is published and when a PT tag is applied. Relying on these tags may thus hinder access to the most up-to-date evidence available. Therefore, we instead use an automated tagging system that uses machine learning to classify articles as RCT reports (or not). This model has been validated extensively in prior work \cite{marshall2018machine}, and we do not describe it further here. 

Next, we replace all abbreviations with their long forms using the Ab3P algorithm \cite{sohn2008abbreviation}. Using long forms has the complementary advantages of improving PICO labeling accuracy while also reducing the amount of context needed for prediction by downstream model components. 

% bcw: sticking w/Ani's suggestion for now
%\subsection{Clinical Concept Recognition}
\subsection{Study Descriptor Recognition}
\subsubsection*{PICO Elements}
In order to identify the spans of text corresponding to the PICO elements of the trial, we use the \emph{EBM-NLP} corpus \cite{nye2018corpus}. This is a dataset comprising $\sim$5,000 abstracts of RCT reports that have been annotated to demarcate textual spans that describe the respective PICO elements. In addition to these spans, it contains more granular annotations on information within spans (e.g., specific Population attributes like age and sex). 

% bcw 4/29: are we sure we want to call them 'entities'? to hedge bets i have at least said 'clinical entities'
\begin{table}[]
\small 
    \centering
    \begin{tabular}{lccc}
         & F1 & Precision & Recall \\ \cline{2-4}
         Tokens & 0.63 & 0.56 & 0.72 \\
         Clinical Entities & 0.67 & 0.55 & 0.87 \\
    \end{tabular}
    \caption{Macro-averaged scores for ICO span prediction at both the token and clinical entity level.}
    \label{tab:pico}
\end{table}

We follow our prior work \cite{nye2018corpus} in training a BiLSTM-CRF model that learns to jointly predict each PICO element using EBM-NLP. Recent work has shown the efficacy of BERT \cite{devlin2018bert} representations in this space, e.g., Beltagy \emph{et al.} achieved state-of-the-art performance on EBM-NLP using this approach \shortcite{beltagy2019scibert}. Therefore, for all text encoding we use BioBERT \cite{DBLP:journals/corr/abs-1901-08746}, which was pre-trained on PubMed documents.\footnote{For PICO tagging on EBM-NLP we found that BioBERT performed comparably to SciBERT \cite{beltagy2019scibert}.}
%which is more closely aligned with the task domain and offers equivalent improvements on the PICO labeling task. 

% bcw 1/28/20: clarify which dataset these results are over -- the newly collected stuff or canonical EBM-NLP? if the latter, dev or test?
Results for Interventions/Comparators and Outcomes on our test set are reported in Table~\ref{tab:pico}. Since these spans will serve as inputs to downstream models in the pipeline, high recall at the expense of precision is preferable; we will allow subsequent classifiers to discard spurious spans. We achieve 0.87 recall at the clinical concept level.

\subsubsection*{Evidence Statements}

In addition to PICO elements, we extract all sentences in the abstract that are predicted to contain evidence concerning the relative efficacy of an Intervention. Our training data for this model is sourced from the \emph{Evidence-Inference} corpus \cite{lehman2019}, which comprises $\sim$10,000 annotated `prompts' across $\sim$2,400 unique full-text articles. Each prompt specifies an Intervention, a Comparator, and an Outcome. Doctors have annotated the prompts for each article, supplying an extracted snippet that presents the conclusion for these ICO elements, as well as an inference concerning whether the Outcome increased, decreased, or remained the same in the intervention group (as compared to the comparator group).

\begin{table}[]
\small 
    \centering
    \begin{tabular}{lccc}
         & F1 & Precision & Recall \\ \cline{2-4}
         Evidence & 0.69 & 0.53 & 0.97 \\
    \end{tabular}
    \caption{Performance for identifying evidence-bearing sentences.}
    \label{tab:ev}
\end{table}

% bcw 2/28/20 -- again specify what test set this is
We frame evidence identification as a sentence classication task, and train a linear classification layer on top of BioBERT outputs. Our positive training examples are the sentences containing evidence snippets in Evidence-Inference, and we draw an equal number of length-matched negatives randomly from the rest of the document. As shown in Table~\ref{tab:ev}, we achieve extremely high recall on the test set, but only middling precision. On inspection, many of these false positives are sentences from the conclusion that provide a high-level summary of the evidence, but aren't the best evidence statement --- as provided by the annotator --- for any given ICO prompt.

\subsection{Relation Extraction}
\label{section:relation-extraction}

To transform the extracted spans into a semantic representation of the trial that can be used to construct an evidence map, we must identify all instances of an outcome being reported, and infer which two treatments were being directly compared as the intervention and comparator with respect to said outcome. Finally, given each assembled ICO prompt, we can then predict the trial's findings regarding whether the outcome increased, decreased, or was not statistically different under the intervention versus the comparator.
In effect, we are aiming to jointly extract ICO prompts \emph{and} infer the directionality of the results reported concerning these, whereas prior work \cite{nye2018corpus,lehman2019} has considered these problems only in isolation.

Our strategy for assembling ICO prompts is informed by the style in which results are commonly described in abstracts.%, %even across different medical domains. 
When results are described in an article the outcome is typically referenced explicitly, while the intervention and especially the comparator are often referenced either indirectly (``Mean headache duration was similar between groups''), or not at all (``No significant difference was observed for recovery time''). In the fully annotated dev set collected for this work, 87\% of outcomes were described explicitly in an evidence span, while only 28\% of treatments were explicit.
%In the Upwork dev set, 87\% of outcomes appear in an evidence span, while only 28\% of treatments do. % bcw 1/28/20: clarify which dev set

Motivated by this observation, we use the (explicit) outcomes extracted from an evidence snippet as a starting point;  for each of these outcomes, the associated intervention and comparator are then inferred.
This has the significant advantage of explicitly linking each outcome to the evidence that will be used to infer the directionality of the reported finding. This also provides the end-user with an interpretable rationale for the inference concerning treatment efficacy.
% bcw: maybe cite ERASER here...

To link candidate extracted treatments to specific outcome mentions, we train a model that takes in a candidate treatment, an evidence statement containing the outcome, and the surrounding context from the document, and predicts if the treatment is the participating intervention, the participating comparator, or if it is not involved in this particular evaluation. We use the evidence-inference corpus to provide training examples for the first two classes, and manually generate negative samples for the final class. The negatives are constructed to mimic common errors that the treatment extraction module made on the dev set, including: mislabeling an outcome as a treatment; extracting compound phrases containing multiple individual treatments; and, finally, extracting spurious spans that don't represent a study descriptor. % bcw 1/28/20: what do we mean by 'random spans'? as in, spurious/false positives maybe?
% bcw 1/28/20: "and manually synthesize negatives for the final class" -- can you clarify here? this was done on the ev-inf corpus?

% 1/27/2020 ben: TODO - remove this table. classification accuracy here is not directly useful. it is only a proxy for how usefull the class probabilities will be for ranking treatment spans 
% 1/31/2020 bcw: I have followed your TODO :)
\begin{comment}
\begin{table}[]
\small 
    \centering
    \begin{tabular}{lccc}
         &  F1 & Precision & Recall \\ \cline{2-4}
         I/C classification & 0.83 & 0.83 & 0.82
    \end{tabular}
    \caption{}
    \label{tab:linking}
\end{table}
\end{comment}

% bcw 1/28/20 -- do we need the 2nd [sep]???
% ben 1/30/20 -- probably doesn't make a difference but the original BERT stuff includes it 
The model is a linear classifier on top of BioBERT. Inputs are constructed as: \textsc{[cls] treatment [sep] evidence. context. [sep]}. We experimented with different slices of the document as the context, and achieved the highest dev performance using the first four sentences of the article. The class probabilities from this model are used to rank the possible interventions and comparators for each outcome, and when sufficiently probable candidates are identified we generate a complete ICO prompt.
%  Results for this model are shown in Table~\ref{tab:linking}.  -- bcw: removed because we dropped this table (see above)

% 1/27/2020 ben: TODO - provide evaluation results for getting the right Os and getting the right I/C pair

\begin{table}[]
\small
    \centering
    \begin{tabular}{lccc}
         &  F1 & Precision & Recall \\ \cline{2-4}
         No Difference & 0.91 & 0.94 & 0.89 \\
         Increased & 0.73 & 0.69 & 0.77 \\
         Decreased & 0.76 & 0.75 & 0.78 \\
    \end{tabular}
    \caption{Per-class prediction scores for each outcome in the test set.}
    \label{tab:conclusion}
\end{table}

After assembling all ICO prompts in a document, we feed them to a final classifier to predict the directionality of findings for each outcome, with respect to the given intervention and comparator. This model is trained over the evidence-inference corpus using the provided I, C, and O spans coupled with the sentences that contain the corresponding evidence statement. Empirically, we found that signal for the classifier is dominated by the outcome text and evidence span, with almost no contribution from the intervention and comparator. This is unsurprising given the regularity of the language used to describe conclusions. The reported directionality of the result is almost exclusively framed with respect to the intervention, and only 4.0\% of all outcomes ever have different results for another I+C linking within the same document. The best performing model input was simply \textsc{[cls] outcome [sep] evidence [sep]}, and the results on the test set are reported in Table~\ref{tab:conclusion}. % bcw 1/28/20: clarify that this was over the dev set?

\subsection{Normalizing PICO Terms}

% bcw: here I think a more extensive evaluation is in order, given the limited eval we've done so far
% ben: issues with the EBM-NLP MeSH labels, WiP

% 1/27/20 ben: cite existing work for biomedical concept normalization?
%              plenty of examples of people doing this for diseases/chemicals/proteins
In order to standardize the language used to categorize the articles with respect to their PICO elements, we turn to the structured vocabulary provided by the National Libaray of Medicine (NLM) in the form of Medical Subject Heading (MeSH) terms. This resource codifies a comprehensive set of medical concepts into an ontology that includes their descriptions, properties, and the structured relationships between them. Each article in the MEDLINE database maintained by the NLM is annotated with the relevant MeSH terms by expert library scientists (subject to the same lag that necessitates an RCT classifier instead of relying on annotated Publication Types).

To induce relevant MeSH terms for an extracted text span, we reproduced the method described in the Metamap Lite paper \cite{demner2017metamap} to extract MeSH terms describing the PICO elements. In short, we generated a large dictionary of synonyms for medical terms algorithmically using data from the UMLS Metathesaurus, with synonyms being matched to unique identifiers pertaining to concepts in the MeSH vocabulary. We used this dictionary to map matching strings in our extracted PICO text to MeSH terms, yielding a set of normalized concepts describing each of the population, intervention, and outcome spans in the documents.

To evaluate the accuracy of this approach, we compare the differences between the MeSH terms produced by our system against those provided by the NLM for the 191 articles that comprise the test set for EBM-NLP.

\begin{table}
\small
    \centering
    \begin{tabular}{lcc}
    \hline
         \textbf{Strict}   & Precision & Recall \\ \cline{2-3}
         Extracted spans & 0.26 & 0.24 \\
         Expert spans & 0.23 & 0.26 \\ \hline
        \textbf{Relaxed}  & Precision & Recall \\ \cline{2-3}
         Extracted spans & 0.32 & 0.34 \\
         Expert spans & 0.31 & 0.34 \\
    \end{tabular}
    \caption{Performance for predicting an article's exact MeSH terms using the rule-base system, run on both the automatically extracted spans and the expert-provided test spans.}
    \label{tab:minimap}
\end{table}

 The test articles are provided with an average of $14.8$ MeSH terms per article, while our system induces $14.0$ terms on average. The strictest evaluation for this module is to require exact matches between the predicted MeSH terms and the official MEDLINE terms -- a daunting task given the $30,000$ possible labels we have to chose from.
 However, because the concepts in the ontology exist in varying levels of specificity (for example \emph{Migraine with Aura} is a subset of \emph{Migraine Disorders}), it is often the case that the predicted MeSH term is sufficiently close to the provided MeSH term for practical purposes, but differs in the level of specificity.
 
 To better characterize the performance of our approach, we therefore also consider relaxing the equivalence criteria to include matching immediate parents or children in the MeSH hierarchy. This modification results in a $42\%$ relative increase in recall and a $23\%$ increase in precision, as shown in Table~\ref{tab:minimap}.

\begin{table}
    \small
    \centering
    \begin{tabular}{p{2.6cm}p{0.5cm}p{2.0cm}p{0.5cm}}
        \multicolumn{2}{c}{False Neg}& \multicolumn{2}{c}{False Pos} \\
        \multicolumn{2}{c}{\textbf{Name  /  Count}}& \multicolumn{2}{c}{\textbf{Name  /  Count}} \\
Humans & 185                  & Patients & 115 \\
Middle Aged & 93              & Aging & 42 \\
Adult & 84                    & Therapeutics & 42 \\
Aged & 62                     & Weights/Measures & 33\\
Double-Blind Method & 50      & Placebos & 33\\
Treatment Outcome & 42        & Time & 21\\
Adolescent & 39               & Serum & 17\\
Prospective Studies & 27      & Safety & 17\\
Time Factors & 20             & Pain & 16\\
Child, Preschool & 20         & Women & 14  \\                    
    \end{tabular}
    \caption{Ten most common over- and under-predicted MeSH terms for the test set of 191 articles.}
    \label{tab:minimap-errors}
\end{table}

We observe that while the absolute accuracy is not high, this technique generally captures the key terms for the PICO elements. The most common mistakes, shown in Table~\ref{tab:minimap-errors}, mostly involve missing age or publication type terms, and systematic differences between the general MeSH terms commonly applied to articles (for example, we might apply \emph{Patients} rather than \emph{Humans}).%, but do apply \emph{Patients}, \emph{Theraputics}, and \emph{Placebos}).
% bcw: what's up with the theraputics and placebos in the last sentence??? -- removed for now? 

A more sophisticated aligment between the way MeSH terms are applied by experts and the terms produced by our system has the potential to improve the overall effectiveness of the tool; we intend to pursue this in future work.

\begin{comment}
\section{Pipeline Results}
One of the primary concerns with assembling pipeline systems comprising many components is how errors propagate and compound through models. To evaluate the overall performance and the failure modes of our system, 
in Table~\ref{tab:conclusion} we enumerate all of the differences between the extracted ICO prompts and the reference prompts. % bcw: is this on test set? in any case shouldn't we tally these error types??
\end{comment}

\section{Illustrative Example}
\label{section:example-usage}

\begin{figure}
\centering
\includegraphics[width=0.5\textwidth]{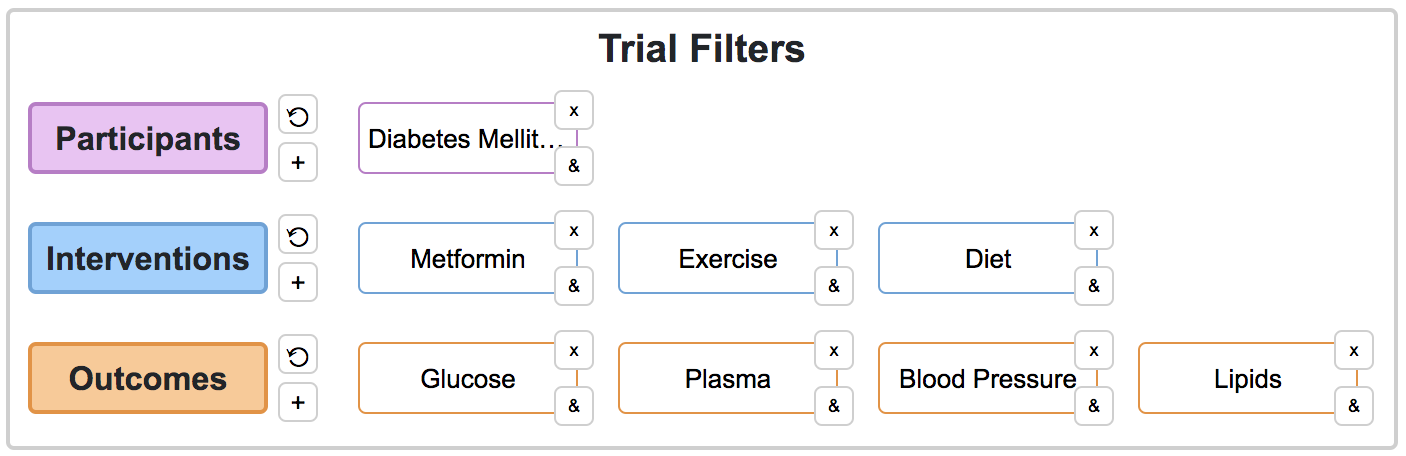} \vspace{-2em}
\caption{View of a collected set of concepts used to specify trials of interest. The search interface allows concepts to be combined using and/or operators.}
\label{fig:pico-terms}
\vspace{-1em}
\end{figure}

To illustrate the envisioned use of our automatic mapping system, we return to the example we began with at the outset of this paper: seeking evidence concerning treatment of Type II Diabetes. 
To begin, the user specifies a condition (Population) of interest. We rely on Medical Subject Headings (MeSH) terms,\footnote{\url{https://www.ncbi.nlm.nih.gov/mesh}} which as discussed above is a structured vocabularly maintained by the NLM. 
We allow users to enter a search string and provide auto-complete options from the MeSH vocabulary. Users can additionally provide interventions or outcomes of interest to further narrow the search. We show an example of a constructed set of filters in Figure~\ref{fig:pico-terms}.

%\begin{figure}
%\centering
%\includegraphics[width=0.5\textwidth]{figures/bar-plot.png} \vspace{-1.5em}
%\caption{Histograms of the most prevalent interventions and outcomes. These counts can be used to iteratively refine search parameters by clicking on bars to add filters, or by hovering over them to view how they correlate with the other elements.}
%\label{fig:bar-plot}
%\end{figure}

Once a set of search terms is specified,   relevant RCTs are retrieved from the comprehensive and up-to-date database.\footnote{We update this database nightly by scanning MEDLINE for new RCT reports using our RCT classifier \cite{marshall2018machine}.} The interface then displays counts of unique interventions and outcomes covered by the retrieved trials.
%This is shown for our running example of diabetes in Figure \ref{fig:bar-plot}.
Each bar in these plots can be clicked to explicitly include that concept in the search terms, allowing for a data-driven approach to building up the search parameters via iterative refinement.

\begin{figure}
\includegraphics[width=0.5\textwidth]{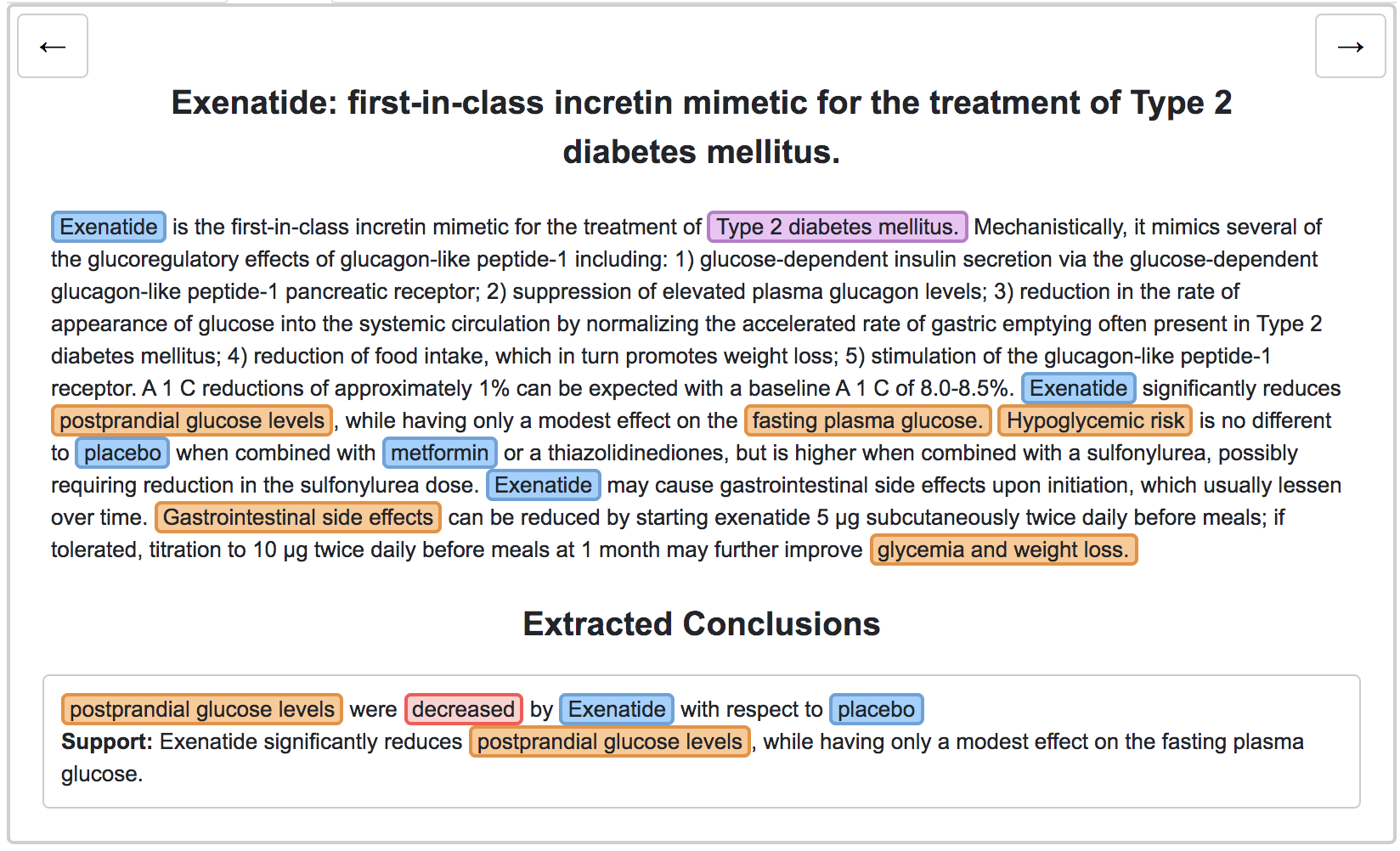}
\vspace{-1.5em}
\caption{Detailed view of selected abstracts that contribute to the evidence map. These are automatically annotated with all extracted information.}
\vspace{-1em}
\label{fig:abstracts}
\end{figure}

At this point, the evidence map shown in Figure~\ref{fig:ev-map} is also displayed, providing a summary of the evidence available for the effectiveness of the selected interventions with respect to their co-occurring outcomes.
The user can mouse-over plot elements to view tooltips that include snippets of contributing evidence from the underlying abstracts, or click through to browse these texts annotated with all of the extracted information, as shown in Figure \ref{fig:abstracts}.

\section{User Study}
To evaluate the system's utility for a real-world task, we provided the tool to a team of researchers at Cures Within Reach for Cancer (CWR4C).\footnote{https://www.cwr4c.org/} Domain experts reviewed the extracted ICO conclusions and automatically generated plots for a randomly selected subset of documents pertaining to cancer trials, a domain that is particularly challenging given the prevalence of complex compound interventions that often share individual components between trial arms.

The reviewers were asked to evaluate the types of mistakes made by the system as well as the overall precision and recall of the extracted conclusions for each document. Across 21 documents average precision was 54\% and average recall was 75\%, and the team expressed excitement about the efficacy of the system for their purposes. CWR4C has continued to work with this tool as a source of information about cancer-related clinical trials.

\section{Conclusions}

We have presented the evidence extraction component of Trialstreamer, an open-source prototype that performs end-to-end identification of published RCT reports, extracts key elements from the texts (intervention and outcomes descriptions), and performs relation extraction between these, i.e., attempts to determine which intervention was reported to work for which outcomes. 

%combines NLP modules with novel relation extraction and classification systems to 
We use this pipeline to provide fast, on-demand overviews of all published evidence pertaining to a condition of interest. 
Moving forward, we hope to refine the linking of extracted snippets to structured vocabularies to run a more comprehensive user-study to evaluate the use of the system in practice by different types of users. 
We also hope to develop a joint extraction and inference model, rather than relying on the current pipelined approach.

\section*{Acknowledgements}

This work was funded in part by the National Institutes of Health (NIH) under the National Library of Medicine (NLM) grant 2R01LM012086, and by the National Science Foundation (NSF) CAREER award 1750978. 
\bibliography{citations}
\bibliographystyle{acl_natbib}

\end{document}